# Relating Chromophoric and Structural Disorder in Conjugated Polymers

Lena Simine, Peter J. Rossky*

Department of Chemistry, Rice University

*Correspondence to: peter.rossky@rice.edu

The optoelectronic properties of amorphous conjugated polymers are sensitive to the details of the conformational disorder and spectroscopy provides the means for structural characterization of the fragments of the chain which interact with light – "chromophores". A faithful interpretation of spectroscopic conformational signatures, however, presents a theoretical challenge. Here we investigate the relationship between the ground state optical gaps, the properties of the excited states, and the structural features of chromophores of a single molecule poly(3-hexyl)-thiophene (P3HT), using quantum-classical atomistic simulations. Our results demonstrate that chromophoric disorder arises through the interplay between excited state delocalization and electron-hole polarization, controlled by the torsional disorder introduced by side chains. Within this conceptual framework, we predict and explain a counter-intuitive spectral behavior of P3HT: a red-shifted absorption, despite shortening of chromophores, with increasing temperature. This discussion expands the conventional view of optoelectronic properties of amorphous materials, by introducing the concept of disorder-induced separation of charges, and by emphasizing the critical importance of optically inert and non-polar side-chains in forcing disorder on the low energy bare structure.

As mentioned above, chromophores are fragments of a molecule or a material which interact with light. For example, in biological polymers such as the photosynthetic complex or fluorescent proteins, the chromophores are distinct molecular units incorporated into a sequence of amino-acids. Their spatial positions within the structure are fixed, and the optical properties are determined by the immediate electrostatic environment. In contrast, in conjugated polymers the chromophore emerges as a fragment of the chain over which the excitation is localized; its position within the polymer, size, associated optical gap, and spatial orientation vary with conformational fluctuations. The static and dynamic variability of these properties constitutes the chromophoric disorder. The most scrutinized aspect of chromophoric disorder, the energetic heterogeneity, is manifest in the width of the absorption/photo-luminescence spectrum and may be interrogated with spectroscopy *(1,2)*. Recently, it has been demonstrated to be structural in origin at the level of a single molecule *(3-5)*. A strong spectroscopic signature of distinct conformational species in low temperature poly(3-hexyl)-thiophene (P3HT) aggregates was reported in Ref. *(6)*. Structural fluctuations in P3HT were captured by spectral diffusion in transient hole-burning experiments in Ref. *(7)*, and the disorder controlled by oligomer length has been investigated using both single molecule and ensemble spectroscopy in Ref. *(8)*. While experimental spectroscopy is the most direct way to interrogate the optical and morphological characteristics of materials, theoretical models can provide a direct molecular interpretation, if the model molecular structures are paired with directly calculated experimental observables.

Here, we use quantum-classical simulations to investigate the effects of structural disorder on the optical properties and the properties of the excited states of our model system P3HT (Fig. 1a). Our choice of the model system is motivated by the fact that alkyl side-chains, attached to the thiophene monomers for increased solubility, cause a modest twist to the nominally planar polythiophene backbone. The degree of twisting modulates the strength of the inter-ring electronic couplings, and affects directly the optoelectronic properties of the polymer *(9-15)*, making P3HT a convenient test-bed for structure-optoelectronic properties relationships. By tuning access to the strained conformations, using temperature as a proxy for more general perturbations, we explore how the various measures of torsional disorder impact optoelectronic properties and, in particular, affect the interplay between spatial de-localization and electron-hole polarization of excited states. Within this conceptual framework, we interpret the trend predicted by our simulations of a single molecule P3HT: a counter-intuitive combination of simultaneous red-shifted absorption and shortened chromophores with increasing temperature, summarized schematically in Fig. 1b; the data is shown in Fig. 2b-d.

To this end, we used a modified Quantum Mechanical Consistent Force Field for π-electrons (QCFF/PI) method *(16, 17)*, with thermalizing stochastic Langevin dynamics to generate thermal ensembles of 30-(3-hexyl) thiophene oligomers (P3HT) across a range of temperatures. The sampling was performed from configurations chosen at 1ps intervals from multiple trajectories, each seeded with uncorrelated initial conditions from a separate trajectory. The trajectories were initiated with extended backbone conformations and were sufficiently short (~100ps) to avoid the complication to spectrum from collapse and self-interaction of the polymer. The absorption spectra were computed using the semi-empirical Pariser-Parr-Pople Hamiltonian for π-electrons at the level of configuration interaction singles (CIS), with a non-polar material environment modelled in the calculation through the alkane-like dielectric constant $\varepsilon=2.7$. Since the nuclear motion is described classically, vibronic spectral features are not captured by this mixed quantum-classical approach. The properties of the exciton, i.e., the centers of mass of the electron, and separately, of the hole, and their localization lengths were derived using a reduced description starting from the CIS mixed density matrix. Following Barford et al. *(18)*, the electron/hole localization length $L_{e/h}$ was defined as the "radius of gyration" of the exciton wave function. For a full discussion of the computational methodology, see Supplementary Methods.

Formally, the chromophore is described by the location and the spatial extent of the excited states with optical gaps within the energy window of the incident radiation. This definition may be simplified by identifying a smaller subset of states of each chromophore that dominate light absorption. In order to identify the relevant states, we calculated the absorption spectrum of a single P3HT molecule, Fig. 2a. In the absence of single molecule absorption experiments, we seek validation from emission studies. A recent single-molecule emission experiment *(5)*, in which Raithel et al. report emission spectrum of single P3HT oligomers trapped in a flash frozen rigid solvent matrix in high temperature conformation, partially validates our model: At the extremely low temperature of the experiment (1.5K) the Stoke's shift is suppressed, and the measured zero phonon line ~530nm agrees well with the peak wavelength produced by our model ~510nm. As shown in Fig. 2a, much of the visible spectrum is captured by the excitation into $S_1$ (green line). Thus, the location and spatial extent of $S_1$ is sufficient to define a chromophore. In agreement with previously proposed definitions, it is the lowest energy instance of so-called Local Exciton Ground States (LEGS) *(19, 20)*.

The sensitivity of the torsional disorder to temperature is evident in Fig. 3a, which shows the torsional potential of mean force (PMF). The PMF, a free energy, is related to the distribution of torsional dihedral angles P(ϕ) through $\text{PMF}(\phi) = -k_B T \log P(\phi)$, where $k_B$ is the Boltzmann constant, and *T* is temperature. It shows two locally stable relative configurations of adjacent thiophene rings: *cis*, with both sulfurs pointing in the same direction ($\phi < 90°$), and *trans* with sulfurs pointing in the opposite directions ($\phi > 90°$). At low temperatures, 10K-100K, the polymer explores the quasi-harmonic minimum of the torsional free-energy potential at ~150° (*trans*) and a much less populated ~45° (*cis*), with very rare deviations. Beyond 100K, the free energy is asymmetrically widened, introducing two effects simultaneously: broadening and deviation from a Gaussian distribution P(ϕ). Ensembles characterized by this non-Gaussian torsional disorder discussed in this work clearly have a "fat-tail" extending into $\phi > 160°$ region. In the case of a single oligomer, discussed here, this *planarization* in the *higher* temperature state is a result of the balance of forces between the steric interactions due to side chains and the forces exerted by the π-electrons on the nuclei, and it reflects an anharmonicity of the torsional potential. In aggregates, similar anharmonicity may be induced by inter-chain interactions. Our single-molecule results apply to such scenarios given that electronic couplings between neighbor chains are small relative to the intrachain couplings *(21-23)*. In our simulations, planar access is gained gradually, with $\phi > 160°$ becoming more populated as the temperature is increased beyond 100K. At temperatures exceeding 200K, the near-planar dihedrals (larger than ~160°) are fully explored by the polymer. In addition to asymmetry, increasing temperature broadens the $\phi$ distribution with the expected increase in π-conjugation-breaking staggered conformations with $\phi \sim 90°$. Therefore, varying the temperature, indeed, allows us to explore the electronic properties of P3HT in a small zoo of distinct conformational species characterized by the $\phi$ distributions of varying width and symmetry.

In polymers such as unsubstituted polythiophene, staggered torsions mark the boundaries of the chromophores, and provide an intuitive illustration of the mechanism of the phononic localization of excitons *(24, 25)*. Exceptions appear in more flexible and disordered polymers, such as P3HT, in which the conjugation is not maximal even in the relaxed conformational ground state; e.g., Coulombic coupling has been shown to extend the excited states over breaks in conjugation *(17, 26)*. The effect of the shape of the torsional free energy on the electronic properties of the P3HT is striking: the distributions of the (re-normalized) inter-ring electronic π-overlap integrals ($\frac{\beta}{\beta_0} = |\cos\phi|$) are shown for low and high temperatures in Fig. 3B and 3C, respectively. At high temperature the distribution peaks at full conjugation, and decays rapidly. This reflects the fact that the anharmonicity of the PMF allows the chromophore to be nearly-flat. At the same time, the tail of the distribution extends to vanishing electronic overlap between adjacent rings, in agreement with the fact that, at high temperatures, highly staggered, conjugation-breaking, torsions are common; this type of disorder gives rise to the chromophores considered in *(9)*. In contrast, at low temperature the distributions are narrower, and they peak at a reduced value of the electronic couplings, implying a weakened, but more uniform conjugation throughout the polymer. The optical properties of P3HT in various states of disorder are summarized in Fig. 2. Figure 2b displays a notable trend: the high temperature ensembles absorb light at longer wavelengths, than the low temperature ensembles. The red-shift is made unambiguous in Fig. 2d, which shows that the average energy of the optical transition to $S_1$ is a decreasing function of temperature.

As shown in Fig. 2c, counter to the usual logic, which suggests that shorter chromophores absorb shorter wavelengths than longer chromophores, the mean conjugation length $L_{e/h}$, i.e., the mean size of the chromophore, is also decreased with heating. The counter-intuitive observation that a red-shifted absorption is correlated with shortened chromophores at higher temperatures can be readily understood in structural terms as follows: On the one hand, the conjugation-breaking torsional twists, accessible at high temperatures, limit the length of chromophores. On the other hand, the nearest neighbor inter-ring electronic coupling (inter-ring electronic overlap) is increased, as shown in Fig. 3c. That is, hot chromophores are shorter but more planar than their cool counterparts. At the same time, the cold chromophores, are more uniformly, but relatively gently, twisted; the twists are ∼ 30°, and conjugation is weaker but persistent. In these conditions, the chromophores occupy longer stretches of the backbone, yet absorb shorter wavelengths on average. This goes against the common epitome for the spectral behavior of conjugated polymers, based on a mental image of an ideal planar conformational ground state, which would imply a blue shift at elevated temperatures. Notably, the appearance of a red shift with increasing disorder has been seen previously in a quite different context, correlated with overall contour bending in single phenylene-vinylene oligomers *(3)*. Next, we show that, at least in the present case, this trend is correlated with a change in electron-hole polarization.

We have seen that ensemble averages of localization lengths and optical gaps showed a strong correlation with structural features of the polymer (Fig. 2c,d). Such clear correlation between oligomer optical transition energies and structural features/conjugation lengths is, however, not evident on a configuration-by-configuration basis. This points to the existence of a hidden process, an additional pathway for energy re-distribution, namely, electron-hole polarization. In order to capture the polar character of the exciton, we characterize each excited state wave function by a parameter D that distinguishes polarized (D>1) from non-polarized (D=1) states, and demonstrate in Fig. 4c that, polarized chromophores constitute the *majority* throughout the considered temperatures. The polarization parameter D is discussed in detail in the Supplementary Note.

More directly, the electron-hole separation is manifest by the readily evaluated displacement between these opposite charge centers. Figures 4a-c summarize the behavior of the electronic polarization, as captured by the mean electron-hole displacement $R_{e-h}$. The monotonic relationships between $R_{e-h}$, the average optical gap $E_{01}$ (Fig. 4a), and the size of the chromophore $L_e$ (Fig. 4b) demonstrate that spectral shifts in disordered materials report the size of the chromophore, as well as its local structural details in a convoluted message. The correct interpretation of this report, here, requires an *a priori* knowledge of either $L_e$ or $R_{e-h}$.

Specifically, Fig. 4a shows that the red-shift of $E_{01}$ is associated with lower electron-hole separation, while the trend in Fig. 4b indicates a larger electron-hole separation for longer chromophores. Torsional disorder is implicit in this data: Longer chromophores typically belong to lower temperature ensembles in which the backbone is more uniformly twisted, while the higher temperature shorter chromophores are, typically, more planar. This suggests that energetic cost of electronic polarization is the price for de-localization over a more disordered fragment of the polymer. This idea is supported by the facts that, first, at low temperatures, when $L_e$'s are longer, a larger fraction of the chromophores is polarized (Fig. 4c), and, second, there is a drastic shift from short to long chromophores in the polar versus non-polar sub-ensembles (compare Fig. S1a-b).

Our results manifest a strong inter-dependence between the key aspects of chromophoric disorder (chromophore sizes, charge-separation character, optical gaps) and conformational disorder in P3HT. This suggests that the tuning of chromophoric disorder may be achieved through the tuning of the torsional potential. Panzer *et al.* have shown that the side-chains, which decorate the polythiophene backbone in P3HT, are instrumental for this process *(6)*. We suggest that control of the electron-hole polarization via morphology is an essential element for the design of organic photovoltaics, through the stability of separated electron-hole pairs and, thereby, the charge transport properties of the material.

## Acknowledgements:

This work was enabled by NSF grant CHE-1641076, and the Extreme Science and Engineering Discovery Environment (XSEDE), which is supported by National Science Foundation grant number ACI-1053575 (35). LS thanks Takuji Adachi, Dominic Raithel, and Richard Hildner for insightful discussions.

## Author Contributions:

P.J.R. supervised the project, L.S. designed and carried out the simulations, and both wrote the manuscript.

## Competing financial interests

The authors declare no competing financial interests.

## References:

[1] Adachi, T., Vogelsang, J., Lupton, J. M. Unraveling the electronic heterogeneity of charge traps in conjugated polymers by single-molecule spectroscopy. *J. Phys. Chem. Lett.* **5**, 573–577 (2014).

[2] Schindler, F., Lupton, J. M., Feldmann, J., Scherf, U. A universal picture of chromophores in pi-conjugated polymers derived from single-molecule spectroscopy. *Proc. Natl. Acad. Sci.U. S. A.***101**,14695−14700 (2004)

[3] Becker, K., Da Como, E., Feldmann, J., Scheliga, F., Thorn Csányi, E., Tretiak, S., Lupton, J.M. How Chromophore Shape Determines the Spectroscopy of Phenylene−Vinylenes: Origin of Spectral Broadening in the Absence of Aggregation. *J. Phys. Chem. B* **112**, 4859-4864 (2008)

[4] Thiessen, A., Vogelsang, J., Adachi, T., Steiner, F., Vanden Bout, D. A., Lupton, J. M. Unraveling the chromophoric disorder of poly(3-hexylthiophene). *Proc. Natl. Acad. Sci. U. S. A.***110**, E3550–E3556 (2013).

[5] Raithel, D., Baderschneider, S., de Queiro, T. B., Lohwasser, R., Köhler, J., Thelakkat, M., Kümmel, S., Hildner, R. Emitting Species of Poly(3-hexylthiophene): From Single, Isolated Chains to Bulk. *Macromol.* **49**, 9553-9560 (2016).

[6] Panzer, F., Sommer, M., Bässler, H., Thelakkat, M., Köhler, A. Spectroscopic signature of two distinct h-aggregate species in poly(3-hexylthiophene). *Macromol.* **48**, 1543–1553 (2015).

[7] Yu, W., Magnanelli, T. J., Zhou, J., Bragg, A. E. Structural heterogeneity in the localized excited states of poly(3-hexylthiophene). *Journal Phys. Chem. B* **120**, 5093–5102 (2016).

[8] Kim, T.-W., Kim, W., Park, K. H., Kim, P., Cho, J.-W., Shimizu, H., Iyoda, M., Kim, D. Chain-length-dependent exciton dynamics in linear oligothiophenes probed using ensemble and single-molecule spectroscopy. *J. Phys. Chem. Lett.* **7**, 452–458 (2016).

[9] Yaliraki, S. N., Silbey, R. J. Conformational disorder of conjugated polymers:Implications for optical properties. *J. Chem. Phys*. **104** (1996).

[10] Brédas, J. L., Street, G. B., Thémans, B., André J. M. Organic polymers based on aromatic rings (polyparaphenylene, polypyrrole, polythiophene): Evolution of the electronic properties as a function of the torsion angle between adjacent rings. *J. Chem. Phys.* **83** (1985).

[11] DuBay, K. H., Hall, M. L., Hughes, T. F., Wu, C., Reichman, D. R., Friesner, R. A. Accurate force field development for modeling conjugated polymers. *J. Chem. Theory Comput.* **8**, 4556–4569 (2012).

[12] Grimm, S., Tabatabai, A., Scherer, A., Michaelis, J., Frank, I. Chromophore localization in conjugated polymers: Molecular dynamics simulation. *J. Phys. Chem. B* **111**, 12053–12058 (2007).

[13] Barford, W., Trembath, D. Exciton localization in polymers with static disorder. *Phys. Rev. B* **80,** 165418 (2009).

[14] Barford, W., Bittner, E. R., Ward, A. Exciton dynamics in disordered poly(p-phenylenevinylene). 2. exciton diffusion. *J. Phys. Chem. A* **116**, 10319–10327 (2012).

[15] Westenhoff, S., Beenken, W. J. D., Yartsev, A., Greenham, N. C. Conformational disorder of conjugated polymers. *J. Chem. Phys.* **125,** 154903 (2006).

[16] Warshel, A., Karplus, M. Calculation of ground and excited-state potential surfaces ofconjugated molecules .1. formulation and parametrization. *J. Am. Chem. Soc*. **94**, 5612 (1972).

[17] Lobaugh, J., Rossky, P. J. Solvent and Intramolecular Effects on the Absorption Spectrum of Betaine-30. *J. Phys. Chem. A* **104,** 899-907 (2000).

[18] Barford, W., Lidzey, D. G., Makhov, D. V., Meijer, A. J. H. Exciton localization in disordered poly(3-hexylthiophene). *J. Chem. Phys.* **133**, 044504 (2010).

[19] Makhov, D. V., Barford, W. Local exciton ground states in disordered polymers. *Phys. Rev. B* **81**, 165201 (2010).

[20] Barford, W. Excitons in conjugated polymers: A tale of two particles. *J. Phys. Chem. A* **117**, 2665–2671 (2013).

[21] Granadino-Roldan, J. M., Vukmirovic, N., Fernandez-Gomez, M., Wang, L.-W. The role of disorder on the electronic structure of conjugated polymers. The case of poly-2,5-bis(phenylethynyl)-1,3,4-thiadiazole. *Phys. Chem. Chem. Phys*. **13**, 14500–14509 (2011).

[22] Qin, T., Troisi, A. Relation between structure and electronic properties of amorphous meh-ppv polymers. *J. Am. Chem. Soc*. **135**, 11247–11256 (2013).

[23] Bittner, E. R., Silva, C. Noise-induced quantum coherence drives photo-carrier gener-ation dynamics at polymeric semiconductor heterojunctions. *Nat. Commun.* **5**, 3119 (2014)

[24] Meier, H., Stalmach, U., Kolshorn, H. Effective conjugation length and UV/VIS spectra ofoligomers. *Acta Polym.* **48**, 379–384 (1997).

[25] Nayyar, I. H., Batista, E. R., Tretiak, S., Saxena, A., Smith, D. L., Martin, R. L. Role of geometric distortion and polarization in localizing electronic excitations in conjugated polymers. *J. Chem. Theory Comput*. **9**, 1144–1154 (2013).

[26] Ma, H., Qin, T., Troisi, A. Electronic excited states in amorphous meh-ppv polymers from large-scale first principles calculations. *J. Chem. Theory Comput*. **10**, 1272–1282 (2014).

[27] Plasser, F., Wormit, M., Dreuw, A. New tools for the systematic analysis and visualization of electronic excitations. I. Formalism. *J. Chem. Phys.* **141** (2014).

[28] Towns, J., Cockerill, T., Dahan, M., Foster, I., Gaither, K., Grimshaw, A., Hazlewood, V., Lathrop, S., Lifka, D., Peterson, G. D., Roskies, R., Scott, J. R., Wilkins-Diehr, N. XSEDE: Accelerating scientific discovery. *CS&E* **16**, 62–74 (2014).

[29] Sterpone, F., Rossky, P. J. Molecular modeling and simulation of conjugated polymeroligomers: Ground and excited state chain dynamics of PPV in the gas phase. *J. Phys. Chem. B* **112**, 4983–4993 (2008).

[30] Di Pierro, M., Elber, R., Leimkuhler, B. A stochastic algorithm for the isobaric-isothermal ensemble with Ewald summations for all long range forces. *J. Chem. Theory Comput.* **11**, 5624–5637 (2015).

[31] Matsumoto, M., Nishimura T. Mersenne twister: A 623-dimensionally equidistributed uniform pseudo-random number generator. *ACM Trans. Model. Comput. Simul*. **8**, 3–30 (1998).

[32] Jailaubekov, A. E., Willard, A. P., Tritsch, J. R., Chan, W.-L., Sai, N., Gearba, R., Kaake, L. G., Williams, K. J., Leung, K., Rossky, P. J., Zhu, X-Y. Hot charge-transfer excitons set the time limit for charge separation at donor/acceptor inter-faces in organic photovoltaics. *Nat. Mater.* **12**, 66–73 (2013).

# FIGURES

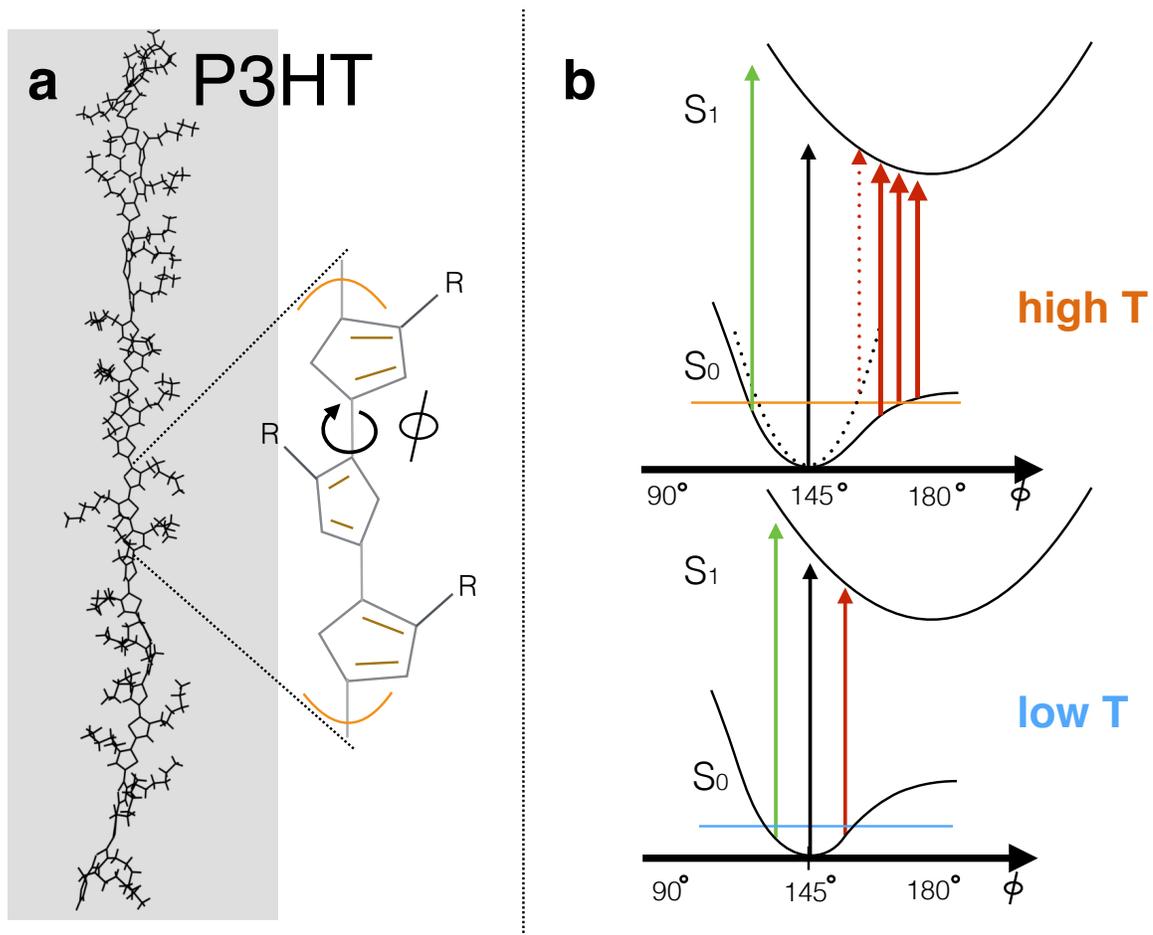

***Fig. 1.*** A schematic representation of P3HT and the mechanism underlying the temperature dependence of the reported bathochromic shift. (a) An illustration of the studied model system: a 30-ring oligomer approximating a single regio-regular P3HT molecule. The inter-ring dihedral angle $\phi$ is defined in the inset. (b) A schematic diagram illustrating the origin of the difference between photo-absorption processes at low and high temperature.

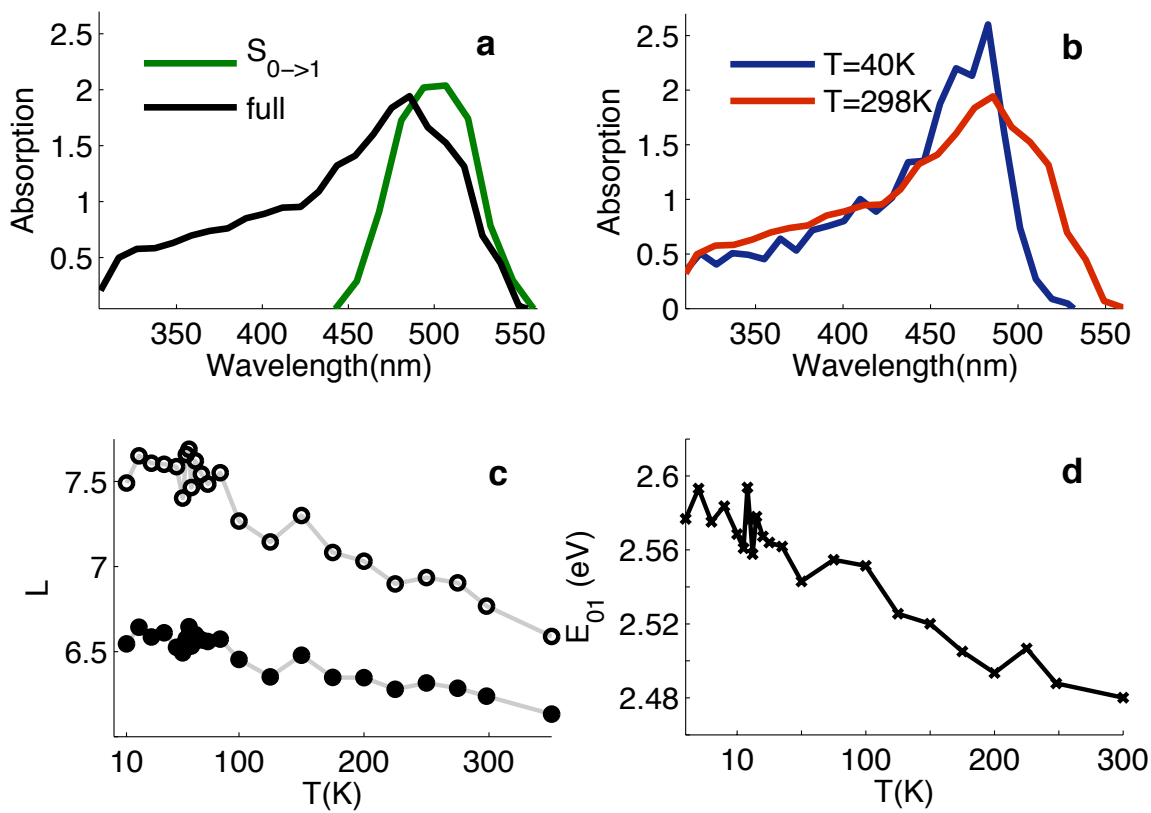

*Fig. 2* The results of the simulations of a single molecule P3HT across a range of temperatures. (a) Full absorption spectrum of the model 30-ring (3-hexyl)thiophene (black); the absorption spectrum of the first excited state (green). (b) Absorption spectra at low temperature (blue), room temperature (red). Note the red-shift of the absorption spectrum at higher temperature. (c) Average localization lengths L (in units of number of rings) of the excited electrons (full circles) and holes (hollow circles) as a function of temperature. The standard error is smaller than 0.1 rings for all points. (d) Average energy gap between the ground and the first excited state, $E_{01}$, as a function of temperature. The standard error is smaller than 0.01eV for all points.

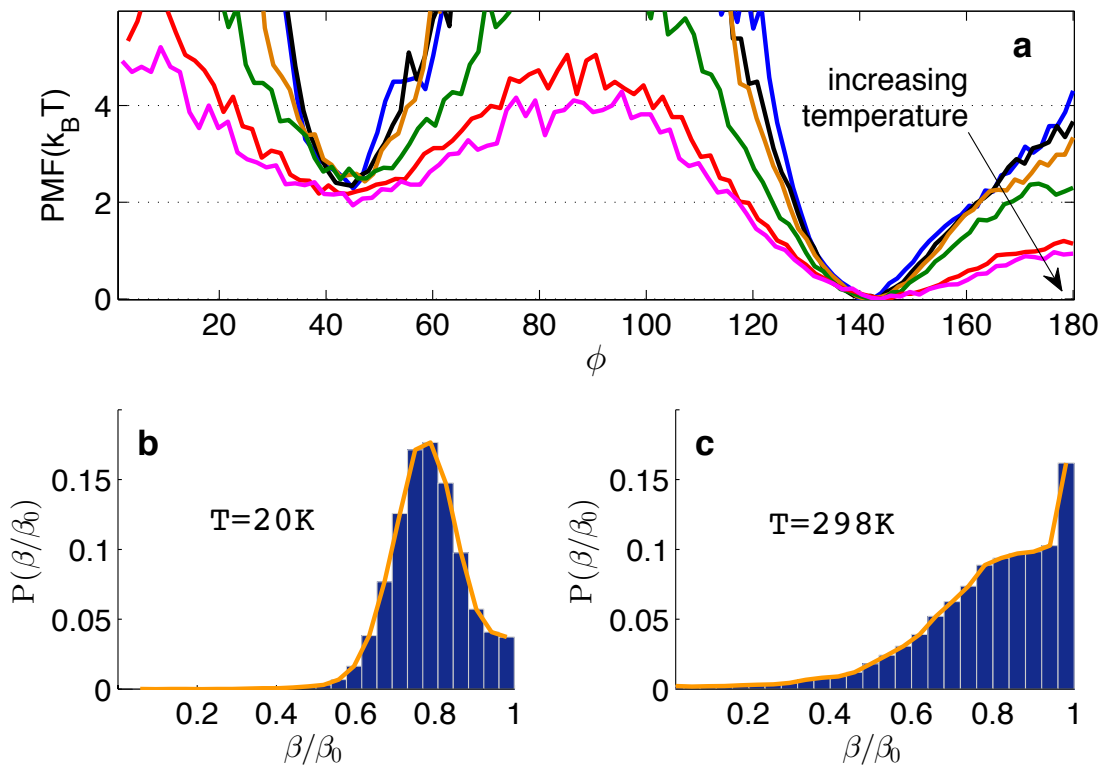

*Fig. 3* The shapes of torsional free energy accessible through temperature and the implications for the electronic properties of the polymer. (a) Potential of mean force for the inter-ring dihedral angle $\phi$, at temperatures 10K, 40K, 60K, 125K, 250K and 298K (units of $k_BT$). Bottom panel: Probability distribution of electronic resonance integrals $\frac{\beta}{\beta_0} = |\cos\phi|$, see equation (S16), sampled at (b) T = 20K and (c) T = 298K.

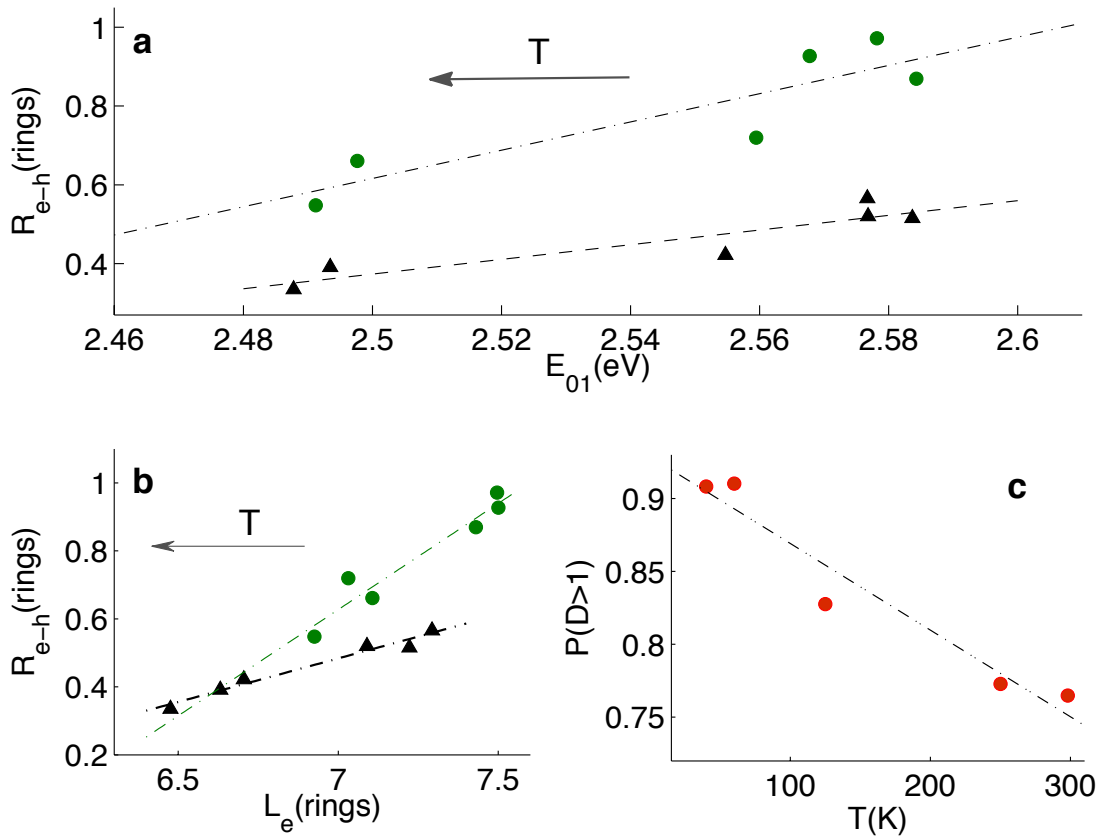

*Fig. 4* The effect of disorder on electron-hole polarization. (a) The linear correspondence between the average electron-hole displacement $R_{e-h}$ and the average of the optical gap $E_{01}$. Each point corresponds to an average taken at a distinct temperature: 10K, 40K, 60K, 125K, 250K and 298K. (b) The fitted linear correspondence between polarization distance $R_{e-h}$ and the localization length $L_e$ at the same temperatures as (a). Black triangles are averages over the entire ensemble, and green points – over the polar (D>1) sub-ensemble. (c) Fraction of polar species (D>1) in the ensemble as a function of temperature. The arrows point in the direction of temperature increase. The fitted lines are provided as a guide to the eye. For the discussion of the polarization parameter D, see Supplementary Note.

# Supplementary Information for

Relating the Chromophoric Disorder and Structural Disorder in Conjugated Polymers

Lena Simine, Peter J. Rossky*

*correspondence to: peter.rossky@rice.edu

Methods

As mentioned in the main text, we use the modified Quantum Mechanical Consistent Force Field for π-electrons (QCFF/PI) method *(16,17)*, with thermalizing stochastic Langevin dynamics to generate thermal ensembles of 30-(3-hexyl)thiophene oligomers (P3HT) across a range of temperatures. The absorption spectra are computed at the level of configuration interaction singles (CIS), and the properties of the exciton, i.e., the centers of mass of the electron, and separately, of the hole, and their localization lengths are derived using a reduced description starting from the CIS mixed density matrix,

$$\rho = \sum_i p_i |\Phi_i\rangle\langle\Phi_i| = \sum_{i,a,r,b,s} c_a^r c_b^{s*} |\Psi_a^r\rangle_i \langle\Psi_b^s|_i \quad (S1)$$

where the index *i* runs over all accessible nuclear configurations in the thermal ensemble, the weight $p_i$ represents the probability of the $i^{th}$ nuclear configuration and the wave-function $|\Phi_i\rangle$ is given by the configuration interaction series with coefficients $c_a^r, c_b^s$ and the Slater determinants $|\Psi_a^r\rangle_i$ for which the indices *a,b* run over the ground state orbitals and *r,s* run over the virtual orbitals.

The reduced density matrix is constructed for the excited electron by tracing out the contributions from all but one relevant molecular orbital and is given in the space of virtual molecular orbitals by,

$$\rho_e = \sum_{a,r,s} c_a^r c_a^{s*} |\phi_r^v\rangle\langle\phi_s^v|. \tag{S2}$$

Analogously, by tracing out all the states occupied by electrons we arrive at the reduced density matrix for the hole,

$$\rho_h = \sum_{a,r} |c_a^r|^2 |\phi_a^{oc}\rangle\langle\phi_a^{oc}|. \tag{S3}$$

The superscripts *v* and *oc* emphasize that the orbital is virtual or occupied.

The mapping from molecular orbital basis to atomic basis is achieved by the usual unitary matrix of molecular orbital coefficients.

The observables (*O*) are computed for the excited electron and for the hole separately as

$\langle O_{e/h}\rangle = \mathrm{Tr}(O\rho_{e/h})$.

Following Barford et al., the electron/hole localization length $L_{e/h}$ is defined as the "radius of gyration" of the exciton wave function *(18)*

$$L_{e/h} = \sqrt{2\langle n^2\rangle_{e/h} - \langle n\rangle_{e/h}^2}, \tag{S4}$$

where n= [1,2...,150] is a discrete coordinate corresponding to the position of the π atoms within the polymer, and the averages are performed using the density matrix $\rho_{e/h}$ as $\langle n_{e/h} \rangle = \text{Tr}(n\rho_{e/h})$ and $\langle n_{e/h} \rangle^2 = \text{Tr}(n^2\rho_{e/h})$

QCFF/PI

Here we provide the minimal details of the method, described previously in Ref. *(16)*, necessary for the discussion. The ground state potential is divided into the classical $V_\sigma$ and quantum $V_\pi$ parts:

$$V_{TOT} = V_\sigma + V_\pi, \tag{S5}$$

The classical part captures the dynamics of the nuclei, while the quantum part represents the π-electrons and includes the interaction between electrons and nuclei via electrostatic potential. The dynamics were integrated using the OVRVO generalization of the Velocity Verlet algorithm *(30)* with the Mersenne Twister pseudo-random number generator *(31)*. At each temperature, the ensemble was constructed by picking configurations from multiple 100ps-long trajectories at 1ps intervals. The trajectories were seeded with uncorrelated initial conditions, taken from a separate room temperature trajectory, and subsequently cooled down, or heated up to the target temperature. Because of the short duration of the individual trajectories, the polymers were never observed to collapse upon themselves, allowing us to explore the optical properties of the experimentally relevant extended conformations.

Molecular Mechanics Force Field

An empirical force field describes the mechanical interactions between all atoms in the simulated system. The classical potential $V_\sigma$ is composed of bonded (B) and non-bonded (NB) interactions

$$V_\sigma = V_{\sigma,conj}^B + V_{\sigma,conj-sat}^B + V_{\sigma,UD}^B + V_\sigma^{NB} \qquad (S6)$$

where the subscript *conj* stands for the subset of atoms in the systems which contribute electrons to the π network, *sat* refers to saturated atoms, which are electronically inert, and *D* stands for the unique dihedral potential for the inter-monomer torsions. The bonded potential between conjugated atoms is given by,

$$\begin{aligned} V_{\sigma,conj}^B &= \sum_i e^{2\alpha(b_i-b_0)} - 2e^{\alpha(b_i-b_0)} \\ &+ \tfrac{1}{2}\sum_i K_\theta(\theta_i - \theta_0)^2 + F(q_i - q_0)^2 \\ &+ \tfrac{1}{2}\sum_i K_\phi^{(1)} \cos\phi_i + K_\phi^{(2)} \cos 2\phi_i \\ &+ \sum_i K_{\theta,\theta'}(\theta_i - \theta_0)(\theta_{i'} - \theta_0)\cos\phi_i \end{aligned} \qquad (S7)$$

where the first term is a Morse potential, which captures the bond stretching between two adjacent π atoms, with $b_i$ is the position coordinate and $b_0$ is an equilibrium bond length; the second term captures the angular bending with $\theta_i$ - the angle defined by three directly bonded atoms and $q_i$ - the distance between the first and the third atoms with the force constants $K_\theta, F$ respectively. The torsional potential in the third term describes the torsional angles $\phi_i$ formed by three consecutive bonds, with one-fold $K_\phi^{(1)}$ and two-fold $K_\phi^{(2)}$ force constants. The last term adds a contribution from the interaction of pairs of two bending angles and the dihedral angle, sharing a common bond.

The bonded potential is given by

$$V^B_{\sigma,conj-sat} = \tfrac{1}{2} \sum_i K_b (b_i - b_0)^2 + 2D_b \tag{S8}$$

$$+ \tfrac{1}{2}\sum_i K_\theta (\theta_i - \theta_0)^2 + F(q_i - q_0)^2$$

$$+ \tfrac{1}{2} K_\phi^{(2)} \cos 2\phi_i,$$

where $K_b, D_b$ are harmonic force constants, and $K_\theta, F$ and $K_\phi^{(2)}$ as before.

Finally, the uniquely defined torsional potential for the torsional angles $\phi_j^T$ between the thiophene monomers is given by

$$V^B_{UD} = \sum_j K_1^{UD} \cos(\phi_j^T) + K_2^{UD} \cos(2\phi_j^T), \tag{S9}$$

with the index $j$ running over all inter-monomer bonds.

The non-bonded term is given by

$$V^{NB}_\sigma = \sum_{ij} A_{ij}\, e^{-\mu_{ij} r_{ij}} - B_{ij} r_{ij}^{-6} \tag{S10}$$

where $r_{ij}$ is the distance between two non-bonded atoms, and parameters $A_{ij}$ and $B_{ij}$ and $\mu_{ij}$ determine the shape of interaction. Most of the parameters for the classical force field were taken from Ref. *(14)* with the exception of the force constants associated with the inter-monomer dihedrals, for which the following values were adopted $K_1^{UD} = 5.725$ and $K_2^{UD} = 0.11$.

Pariser-Parr-Pople (PPP) Hamiltonian

The electronic part of the potential is given by

$$V_\pi = \sum_{\mu\nu} P_{\mu\nu}(H_{\mu\nu} + F_{\mu\nu}), \tag{S11}$$

Where $P_{\mu\nu}$, $H_{\mu\nu}$ and $F_{\mu\nu}$ are the elements of the bond-order matrix **P**, the one-electron Core matrix **H** and the Fock matrix **F** respectively, with the indices $\mu, \nu$ running over the π-atoms. The bond-order matrix

$$P_{\mu\nu} = 2\sum_i c_\mu^i c_\nu^i \tag{S12}$$

represents the electronic overlaps between all pairs of π-atoms, with the index *i* running over the occupied molecular orbitals (MO's), and $c_\mu^i, c_\nu^i$ are the molecular orbital coefficients at atoms $\mu$ and $\nu$. Fock matrix elements are described by,

$$F_{\mu\nu} = H_{\mu\nu} - \tfrac{1}{2} P_{\mu\nu} \gamma_{\mu.\nu} \tag{S13}$$

$$F_{\mu\mu} = H_{\mu\mu} + \tfrac{1}{2} P_{\mu\mu} \gamma_{\mu.\nu} + \sum_{\rho \neq \mu} P_{\rho\rho} \gamma_{\mu.\rho},$$

where the electron-electron Coulomb repulsion matrix elements $\gamma_{\mu\nu}$ are given by

$$\gamma_{\mu,\mu} = \gamma^0_{\mu,\mu} + G_s e^{-2\mu_\beta \left(R_{\mu,\mu+1}-R^{eq}_{\mu,\mu+1}\right)} \cos^2(\phi_{\mu,\mu+1}) \qquad (S14)$$

$$+ G_s e^{-2\mu_\beta \left(R_{\mu,\mu-1}-R^{eq}_{\mu,\mu-1}\right)} \cos^2(\phi_{\mu,\mu-1})$$

$$\gamma_{\mu,\mu+1} = \frac{e^2}{\varepsilon(R_{\mu,\mu\pm 1}+a_{\mu,\mu\pm 1})} - G_s e^{-2\mu_\beta \left(R_{\mu,\mu\pm 1}-R^{eq}_{\mu,\mu\pm 1}\right)} \cos^2(\phi_{\mu,\mu\pm 1})$$

$$\gamma_{\mu,\mu+m} = \frac{e^2}{\varepsilon(R_{\mu,\mu\pm m}+a_{\mu,\mu\pm m})}, m > 1$$

with

$$a_{\mu,\nu} = \frac{e^2}{\gamma^0_{\mu,\mu}+\gamma^0_{\nu,\nu}}. \qquad (S15)$$

In the above, $e$ is the charge of the electron, $R_{\mu\nu}$ is the inter-atomic distance between two π-atoms and $\phi_{\mu,\nu}$ is the torsional angle used to parametrize the disruption in π conjugation between π orbitals of atoms μ and υ due to misalignment of π orbitals for a particular arrangement of the nuclei. Furthermore, $G_s$ is the Slater-orbital overlap matrix parameter *(14)* re-normalized by the dielectric constant ε= 2.7; $\gamma^0_{\mu,\mu} = \frac{I-A}{\varepsilon}$ is calculated as the ratio of the difference between the valence ionization potential *I* and the electron affinity *A*, and the dielectric constant of the medium, ε.

The off-diagonal elements of the one-electron core matrix **H** are treated at the level of Hückel theory

$$H_{\mu\nu} = \beta^0_{\mu\nu} \cos \phi_{\mu,\nu}, \qquad (S16)$$

where μ, υ are restricted to the nearest neighbors and $\beta^0_{\mu\nu}$ is the electronic resonance parameter. The Hückel parameters at perfect conjugation are given by

$$\beta^0_{\mu\nu} = \frac{\hbar^2}{m_e R_{\mu\nu}} \frac{dS_{\mu\nu}}{dR_{\mu\nu}}, \qquad (S17)$$

where ℏ is the reduced Planck's constant, $m_e$ is the electron mass, and $S_{\mu\nu}$ is the overlap integral of the $p_z$ orbitals for atoms $\mu$ and $\nu$. These are approximated by the Linderberg expression

$$\beta^0_{\mu\nu} = e^{-\zeta_{\mu\nu}(R_{\mu\nu}-R^{eq}_{\mu\nu})}\left(\beta^{\mu\nu}_1 + \beta^{\mu\nu}_2(R_{\mu\nu} - R^{eq}_{\mu\nu})\right), \qquad (S18)$$

fitting Eqn. S16 using $\zeta_{\mu\nu}$, $\beta^{\mu\nu}_1$ and $\beta^{\mu\nu}_2$ parameters.

The diagonal elements of **H** are given by

$$H_{\mu\mu} = \alpha_\mu - \sum_{\rho \neq \mu} Z_\rho \gamma_{\mu\rho}, \qquad (S19)$$

Where $\alpha_\mu$ represents the valence ionization parameter at the atom μ, while the electron-core repulsion is given by the second term, with $Z_\rho$ - the nuclear charge on atom ρ (equal to the number of π-electrons this atom contributes). The MO coefficients then follow from the self-consistent field calculation.

The parametrization of the PPP Hamiltonian was adopted from previous work by Rossky group, for a complete list of parameters see the supplementary materials in Ref. *(29, 32)* with the exception of S (Sulfur) for which the following parameters were used: $\gamma^0_{\mu\mu} = \frac{9.79}{\varepsilon}$ $eV$, $\alpha^0_\mu = -20.0$ $eV$, $G_s = \frac{8.16}{\varepsilon}$, $\mu_\beta = \zeta_{CS} = 1.54$ $Å^{-1}$, $R^{eq}_{CS} = 1.41Å$, $\beta^{CS}_1 = -2.12$ $eV$, $\beta^{CS}_2 = 1.31 \frac{eV}{Å}$.

Note

Within the CIS representation, we may divide the ensemble into polarized and non-polarized chromophores, using a polarization index D, which we define as the number of excitations that contribute significantly to $S_1$ (CIS coefficients greater than 0.1). Thereby, the sub-ensemble of polarized chromophores includes all species with D>1, while D=1 species are considered to be non-polar. We stress that D captures only the nature of the molecular orbitals, not correlation, and, more specifically, the asymmetry between the orbitals describing the hole and those describing the excited electron. While more sophisticated approaches exist for the analysis of electronic excitations, e.g. *(27)*, we leave methodological refinement for future work.

In our simple description, the electron-hole polarization is manifest by the increased relative displacement between the opposite charges. The comparison of Fig. S1c and S1d shows that chromophores with D>1, have a broader distribution of electron-hole displacements, $r_{e-h}$, than the sub-ensemble with D=1. Furthermore, the distribution of the localization lengths of the electrons $l_e$ in the D>1 sub-ensemble is shifted to longer lengths with respect to the non-polarized distribution, see Fig. S1a and S1b ($l_h$, the localization lengths of the holes, follow a similar trend). The distributions of absorption optical gaps $E_{01}$ for both sub-ensembles are shown in Fig. S1e and S1f: the non-polar chromophores (D=1), which are shorter, and give rise to slightly red-shifted absorption energies, relative to the polar sub-ensemble (D>1).

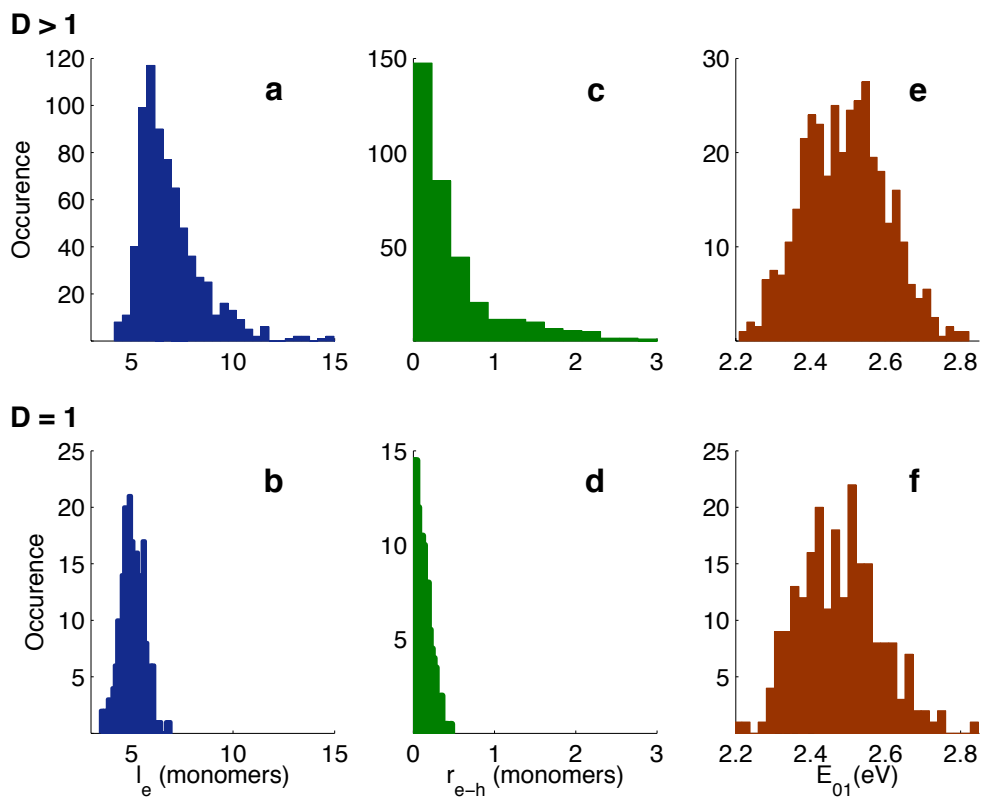

*Fig. S1* Distributions of sizes of the chromophores $l_e$, electron-hole separation distance $r_{e-h}$ and $S_{0\rightarrow1}$ transition energy $E_{01}$. (**a,c,e**): sub-ensemble with the value of polarization parameter D>1; (**b,d,f**): sub-ensemble with D=1 at low T.